\documentclass[10pt,journal]{IEEEtran}
\usepackage[english]{babel}
\usepackage[combine]{ucs}
\usepackage{url}
\usepackage{amsmath}
\usepackage{changepage}
\usepackage[colorinlistoftodos]{todonotes}
\usepackage{color,soul}
\usepackage{threeparttable}
\usepackage{setspace}
\usepackage{booktabs}
\usepackage{fancyhdr}
\usepackage{longtable}
\usepackage{multicol}
\usepackage{enumitem}
\usepackage[sort&compress,numbers]{natbib}
\usepackage{blindtext}
\usepackage{scrextend}
\usepackage{caption}
\usepackage[FIGBOTCAP]{subfigure} 
\usepackage{graphicx}
\graphicspath{{./}}
\begin{document}
\title{Nano-Intrinsic True Random Number Generation}
\author{Jeeson~Kim,~\IEEEmembership{Student Member,~IEEE,} 
Taimur~Ahmed, Hussein~Nili, Nhan~Duy~Truong, Jiawei~Yang, Doo~Seok~Jeong,
Sharath~Sriram,~\IEEEmembership{Member,~IEEE,}
Damith~C.~Ranasinghe,~\IEEEmembership{Member,~IEEE,}
and~Omid~Kavehei,~\IEEEmembership{Senior Member,~IEEE}
\thanks{Manuscript received X; revised X. This work was supported by Australian Research Council under grant DP140103448 and National Natural Science Foundation of China under Grant 61501332.}
\thanks{J.~Kim, N.D.~Truong and O.~Kavehei are with Nanoelectronic and Neuro-inspired Research Laboratory, RMIT~University, VIC~3000, Australia.\protect\\ E-mails: jeeson.kim@rmit.edu.au.}
\thanks{T.~Ahmed, S.~Sriram, J.~Kim and O.~Kavehei are with the Functional Materials and Microsystems Research Group, RMIT~University, VIC~3000, Australia.}
\thanks{H.~Nili is with the Department of Electrical and Computer Engineering, University of California, Santa Barbara, CA~93106~USA.}
\thanks{J.~Yang is with Wenzhou Medical University, China.}
\thanks{D.~S.~Jeong is with the Electronic Materials Research Centre, Korea Institute of Science and Technology, 136-791 Seoul, Republic of Korea.}
\thanks{D.~C.~Ranasinghe and J.~Kim are with the Auto-ID Labs, School of Computer Science, The University of Adelaide, SA~5005, Australia.}}

\maketitle
\begin{abstract} 
Recent advances in predictive data analytics and ever growing digitalization and connectivity with explosive expansions in industrial and consumer Internet-of-Things (IoT) has raised significant concerns about security of people\rq{}s identities and data. It has created close to ideal environment for adversaries in terms of the amount of data that could be used for modeling and also greater accessibility for side-channel analysis of security primitives and random number generators. Random number generators (RNGs) are at the core of most security applications. Therefore, a secure and trustworthy source of randomness is required to be found. Here, we present a differential circuit for harvesting one of the most stochastic phenomenon in solid-state physics, random telegraphic noise (RTN), that is designed to demonstrate significantly lower sensitivities to other sources of noises, radiation and temperature fluctuations. We use RTN in amorphous SrTiO$_3$-based resistive memories to evaluate the proposed true random number generator (TRNG). Successful evaluation on conventional true randomness tests (NIST tests) has been shown. Robustness against using predictive machine learning and side-channel attacks have also been demonstrated in comparison with non-differential readouts methods.
\end{abstract}

\section{Introduction}
Generating unpredictable stream of random sequences is crucial for many, if not all, conventional cryptographic primitives such as key cryptography, digital signatures and ciphers. Internet-of-Things (IoT), wireless networks and radio-frequency identification (RFID) are just a few broad, yet sensitive, examples~\cite{mathew20122,yang201416}. These applications are usually extremely limited in their power, cost and area budgets that demands high efficiency. Due to the lack of efficient sources of randomness, {\it pseudo}-random number generators (PRNG) have been used in different applications and have been frequently reported to be predictable, and hence, vulnerable to a range of attacks~\cite{yang2016all,bae20163}.    

True randomness is usually considered an answer to the issue of {\it pseudo}-randomness. In short, it is extremely difficult, if not impossible, to mathematically prove \lq\lq{}true\rq\rq{} randomness. For true random number generators (TRNGs), entropy source is a very important factor. In hardware intrinsic security, the source has to be a physically genuine random phenomenon. There are numerous spatiotemporal phenomena in hardware, specially at deep-micron or nanometer scales, that have been used as sources of randomness, and chaotic systems~\cite{yang2016all,bae20163,cicek2014new,kim201682nw}. In CMOS, randomness and jitter in oscillators output~\cite{amaki2013process,yang201416}, jitter in digital systems~\cite{kuan20140,robson2014truly}, metastability in common mode comparators as well as well-known sampling uncertainty of D flip-flops~\cite{bae20163}, metastability in latch circuits~\cite{tokunaga2008true}, themal noise~\cite{mathew20122}, and edge racing in even-stage inverter rings~\cite{yang2016all} are just a couple of examples. Oscillator-based and metastable TRNGs have the simplest and largest circuits, receptively with oscillator-based TRNGs suffering from reported poor randomness~\cite{yang2015robust}. Entropy sources like thermal noise in filed-effect transistors (FETs) is strong function of temperature and its noise power is relatively weak, therefore, not only harvesting the noise takes considerable efforts but extensive considerations are required to ensure minimum correlation in the output~\cite{bae20163}.

To evaluate randomness, the National Institute of Standards and Technology (NIST) in the U.S. developed a set of standard statistical tests, which provides a reasonably solid verification ground~\cite{niststs}. However, relying solely on NIST\rq{}s tests comes with an increasingly troublesome consequences and more attention should be drawn to desirable characteristics of a source of entropy. For instance, if an adversarial access or manipulation of environmental factors results in extracting information or successful modeling (for example, using machine learning), there is a real risk of vulnerability. Environmental factors can be listed as, but not limited to, noise, radiation and temperature. A TRNG should ideally be insensitive to environmental factors. Noise amplification, temperature (in case of thermal noise), and dependency on process variation are a few factors that could potentially make TRNGs predictable through creating undesirable bias towards 0 or 1 in the output bit stream, therefore, making the system more vulnerable to a range of attacks~\cite{lampert2016robust}. While it is possible to mitigate dependency of TRNGs to environmental factors by choosing stronger entropy sources, we require noise and temperature aware circuits to harvest the entropy with limited bias imposed on the output. For instance, it could be argued that single event upsets are likely to occur in SRAM\rq{}s power-up metastability-based random number generator (RNG) under radiation~\cite{holcomb2009power}.

Random telegraph noise (RTN) has recently attracted a growing attention as probabilistic and relatively strong source of noise~\cite{figliolia2016true}. RTN has been studied in various types of devices including FETs~\cite{puglisi2015microscopic,grasser2012stochastic,mori2012mechanism,dongaonkar2016random}, carbon nanotube transistors (CNTs)~\cite{liu2008correlated} and broad class of resistive switching memories~\cite{choi2014random,ielmini2010resistance,balatti2014voltage,soni2010probing,raghavan2013microscopic,calderoni2014performance}.   Among these technologies, resistive memory devices have been observed to have one of the strongest RTN signals, hence, circuits presented RTN-based TRNGs using nanometer scale resistive memory devices have been shown without the need for amplification~\cite{huang2012contact}. One of the main open challenges of dealing with random telegraph signals (RTSs) for TRNGs is effective extraction of the noise for maximizing randomness in the output and at the same time, minimizing disturbance and systematic bias mainly due to environmental factors discussed earlier.

This paper presents effective harvesting circuit of stochastic RTN in amorphous SrTiO$_3$ ($a$-STO)-based valency change reduction-oxidation (redox) resistive switching memories (VCM-ReRAMs) and implementation of an innovative ReRAM-based TRNG. This paper emphasis that advantages of differential readout operation, like a metastability-based TRNG in Ref.~\cite{holleman20083}, in increasing noise immunity, higher linearity, relative immunity against temperature and radiation (due to microscale sizes) for TRNG applications outweigh potential increase in area compared to single-ended readout approach reported in literature~\cite{chen2016unified,huang2012contact,karam2016security,balatti2016physical,balatti2015true}. We evaluate randomness quality of output bit stream not only based on NIST tests but also using machine learning attacks. We also showed clear superiority of the proposal differential RTN readout circuitry over those reported in the literature.

In the next section, Section~\ref{sec:noise}, we provide an overview of noise and stochastic RTN in our ReRAMs. Section~\ref{sec:fab} briefly describe device material stack and fabrication process. Section~\ref{sec:rng} discusses RTN characteristics in our devices and the significance of hardware-based random number generators (RNGs), including the proposed RTN harvesting circuitry. Section~\ref{sec:eval} reports the proposed TRNG evaluations and immunity against environmental factors and predictive machine learning attacks.

\section{From Random Telegraphic Noise to Random Number Generators}\label{sec:noise}
In this section, we describe the role of noise in electronic devices, RTN characteristics in $a$-STO-based VCM-ReRAM and review crucial requirements of designing a TRNG.

\subsection{Noise}\label{subsec:noise}
Noise is traditionally considered as an unwanted nondeterministic phenomenon that if not suppressed, it corrupts signal and reduces signal-to-noise ratio (SNR). Noise power can be written as

\begin{equation}\frac {\Delta P_{noise}}{\Delta f}=1/f^{\alpha}.\end{equation}

This is called flicker noise or 1/$f$ noise because its noise spectrum obeys the law as reciprocal of the frequency (1/{\it f}$^{\alpha}$), where the exponent $\alpha$ is very close to unity. 

\begin{figure}[ht]
\centering
\subfigure[]{\includegraphics[width=2.9in]{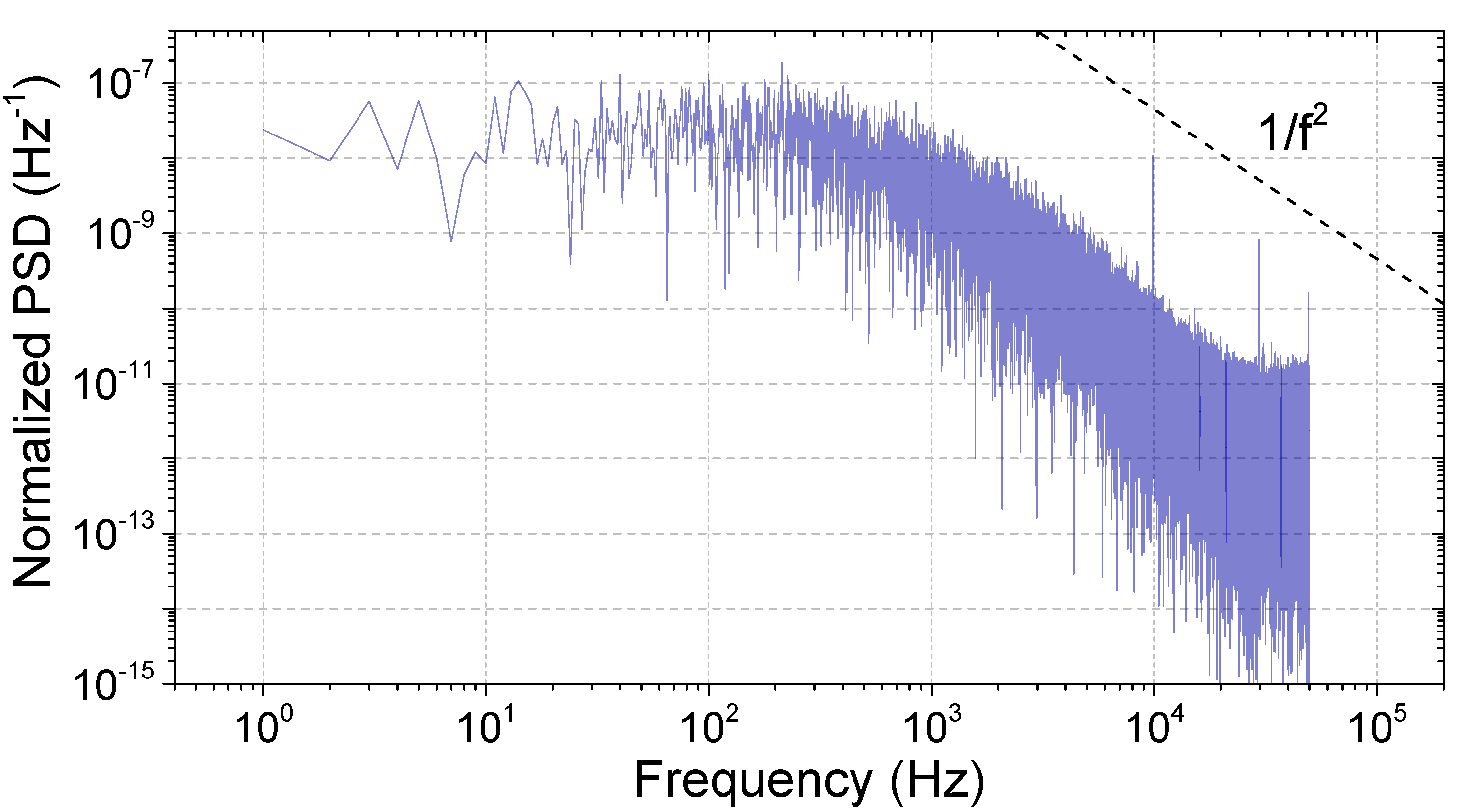}\label{fig:freq}}
\subfigure[]{\includegraphics[width=2.9in]{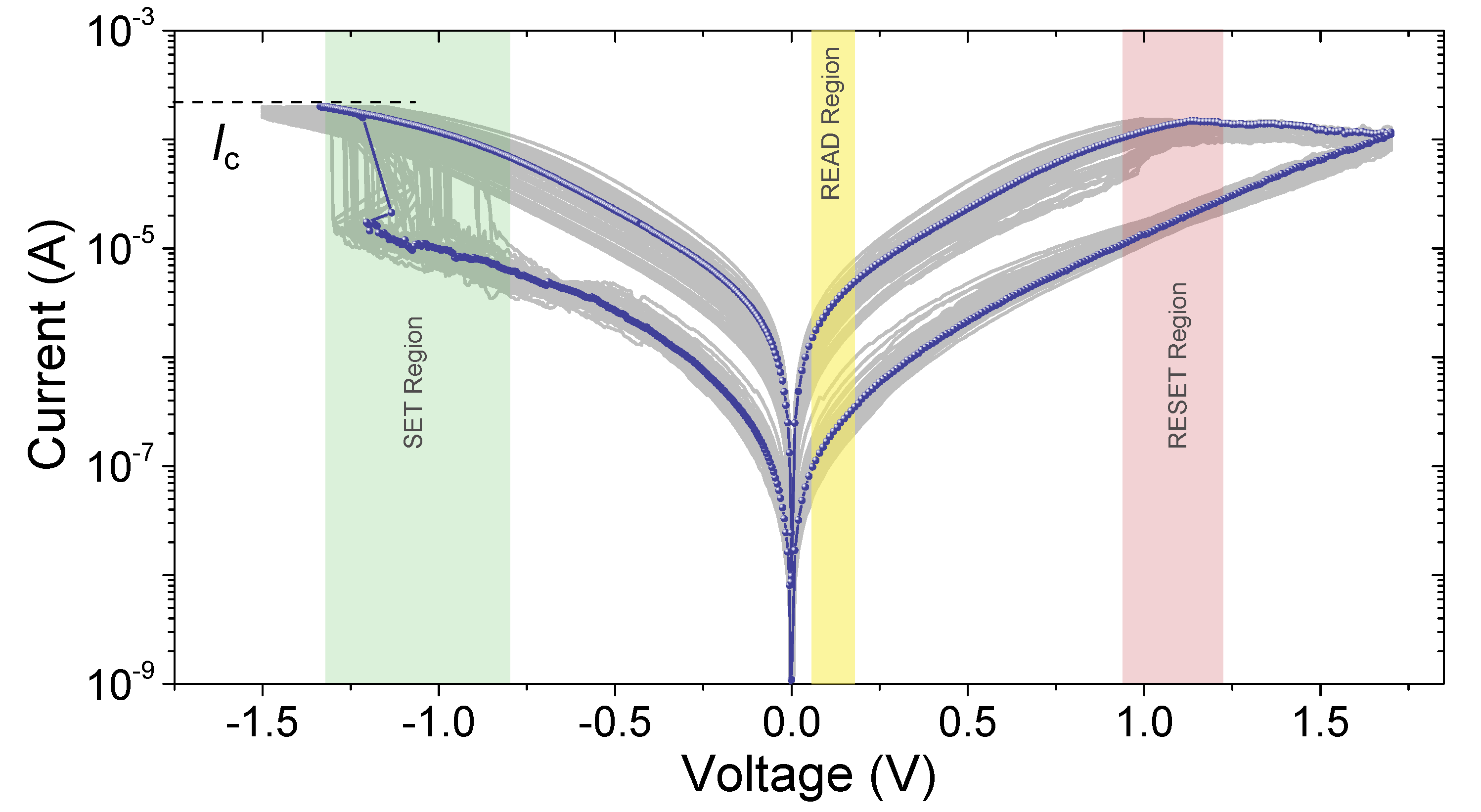}\label{fig:IV}}
\caption{Noise and current-voltage characteristics in ReRAM samples. (a) Normalized PSD measured from one of our $a$-STO-based ReRAM sample programmed at high-resistance state (HRS). ReRAM\rq{}s PDS follows $\sim$1/$f^2$ slope as reported in a number of literature. (b) Current-voltage ($I$--$V$) bipolar-switching signature of fabricated ReRAM. This measurement shows only 50 SET/RESET cycles, with $I_{\rm c}$ representing compliance current set during the measurement. Highlighted voltage ranges show, voltages at which readout operations have been carried out in this paper and voltages at which SET and RESET switchings occur due to cycle-to-cycle and device to device variation.}
\end{figure}

In case of ReRAM, noise data showing $\alpha$ is approaching $2$ for high-resistance state (HRS) as shown in Fig.~\ref{fig:freq} at room temperature, where ReRAM conductance is measured by applying $125$ mV potential across a device, significantly weaker than potentials required to induce enough current into the ReRAM device to impose SET switching (see Fig.~\ref{fig:IV}).

Random telegraphic or \lq\lq{}popcorn\rq\rq{} noise is a random low frequency fluctuation of conductance that appears in two or more current levels. It is believed to be the result of random carrier capture (electron trapped in a local
defect) / emission (empty defect) in/from one or more bistable defects~\cite{ambrogio2014statistical2}. Like many other physical phenomena, it could be described as statistical probability of overcoming a transition barrier from capture to emission or vice versa. Temporal behavior of RTN in ReRAMs is repeatedly shown to be highly random, therefore, it could be used as random source for generating random bits, which is the aim of this work. RTN is often observed in a low frequency regimes of scaled devices and is frequently described as circuit \lq\lq{}designer\rq{}s nightmare\rq\rq{}~\cite{simoen2011invited}. RTN behavior could be described with a number of time constants, shown in this paper with $\tau$.

\begin{figure}[ht]
\centering
\subfigure[]{\includegraphics[width=2.8in,height=1.6in]{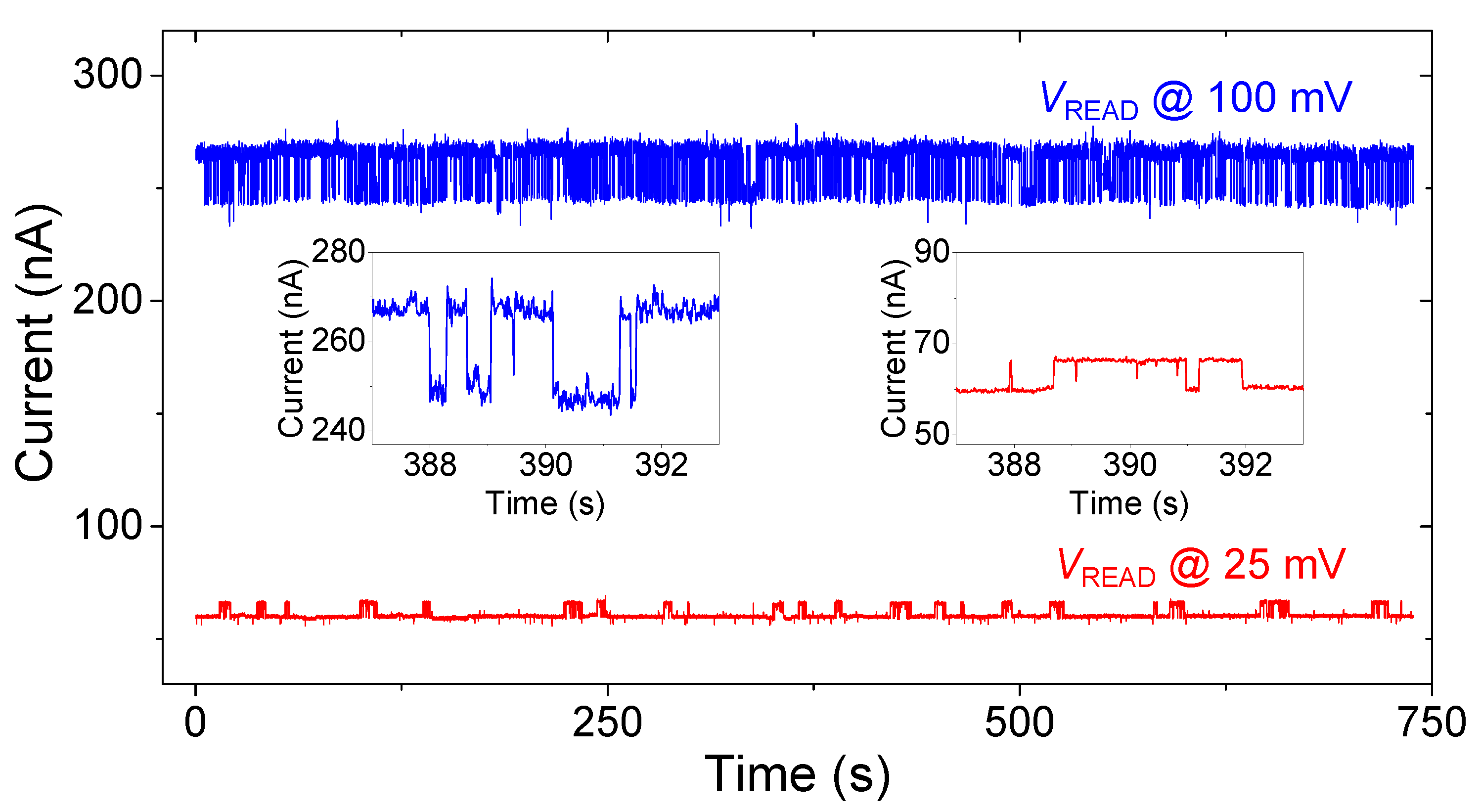}\label{fig:time}}
\subfigure[]{\includegraphics[width=2.8in,height=1.5in]{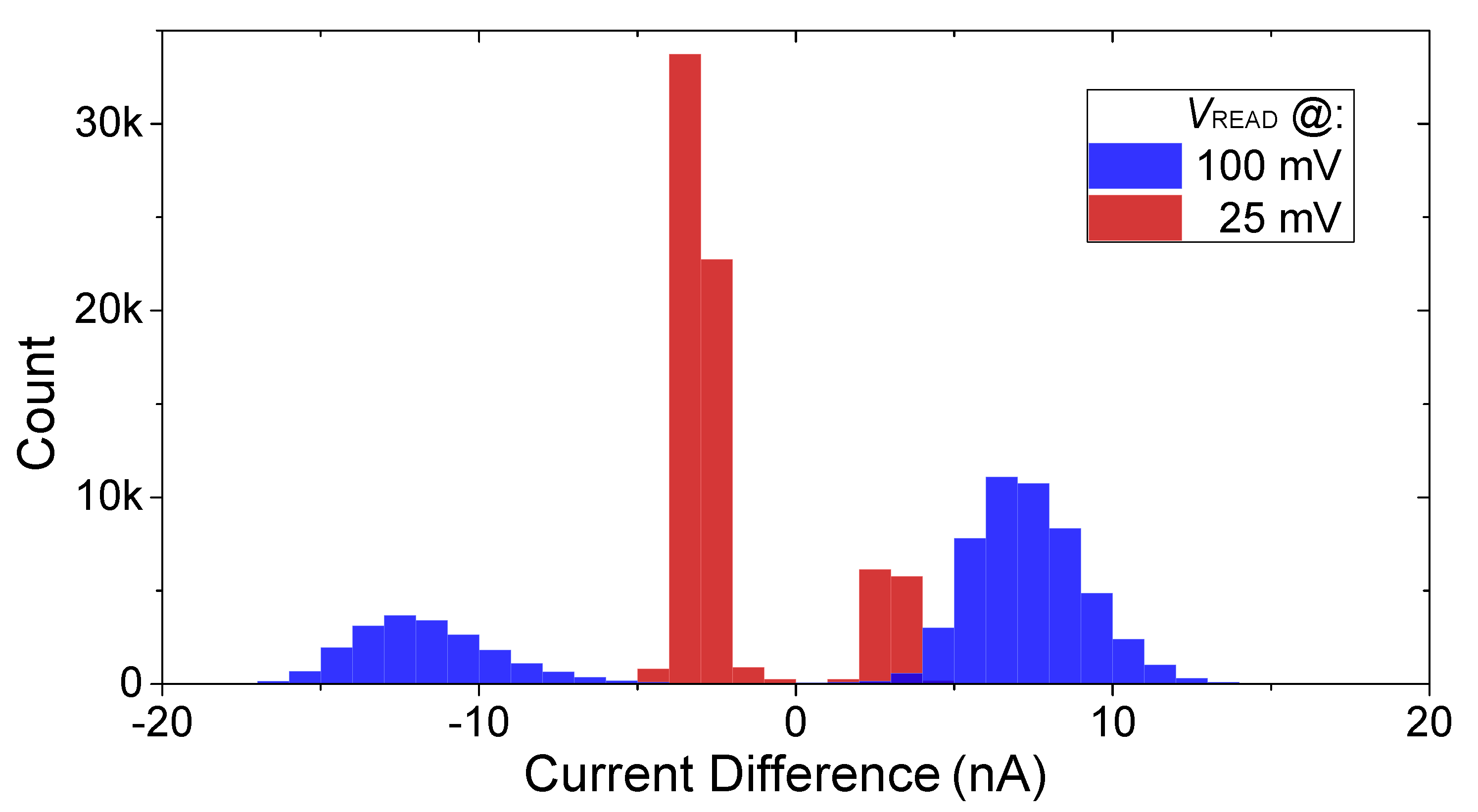}\label{fig:hist}}
\subfigure[]{\includegraphics[width=2.2in]{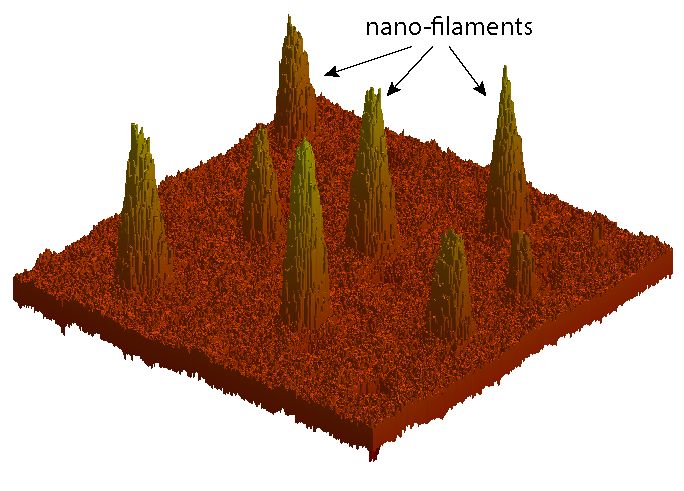}\label{fig:path}}
\caption{ReRAM\rq{}s RTN characteristic. (a) RTN represented in time domain at two READ voltages. Insets show RTN amplitudes. (b) Random telegraphic signal (RTS) amplitude difference between high and low levels of RTN signal. (c) 3D map of conductive nano-filaments extracted from an in-situ scanning probe microscopy (SPM). Defects (traps) are believed to be alongside these filaments.}
\end{figure}

Measured RTN behaviors of a ReRAM at HRS for two different READ voltages at room temperature are shown in Fig.~\ref{fig:time}. It is shown that switching time between different RTN levels (a single trap system here), is a stochastic phenomenon. In frequency domain (Fig.~\ref{fig:freq}), Lorentzian spectrum, 1/$f^2$, starting at corner frequency, $f_{\rm RTS}$. The corner frequency is strong function of $\tau_{\rm L}$ and $\tau_{\rm H}$, which are periods of time a career spent in the low and high levels, respectively (in case of a single trap system here). $f_{\rm RTS}$ can therefore be written as 
\begin{equation}
{f_{\rm RTS}}=\frac{1}{2\pi\tau_{\rm RTS}}=\frac{1}{2\pi}\Big(\frac{1}{\tau_{\rm L}}+\frac{1}{\tau_{\rm H}}\Big).
\end{equation}
PSD then can be calculated by the Wiener-Khintchine formula by taking Fourier transform of the noise-noise autocorrelation~\cite{machlup1954noise}, 
\begin{equation}
S_{\rm RTS}(f)=\frac{4\Delta{I}^2}{\tau_{\rm H}+\tau_{\rm L}} \frac{{\tau_{\rm RTS}}^2}{1+(2\pi f \tau_{\rm RTS})^2},
\end{equation}
where $\Delta I$ represents amplitude of a random telegraphic pulse.

Fig.~\ref{fig:path} illustrates 3D map of an in-situ scanning probe microscopy of our device, highlighting nano-filaments. One or multiple defects/traps alongside these nano-filaments are commonly believed to be the origin of RTN in ReRAM devices. In addition to probabilistic nature of these capture and emission in/from these defects, creation and rupture of these nano-filaments are also a probabilistic phenomenon causing considerable cycle-to-cycle (programming) conductance fluctuation~\cite{na2016offset}. Therefore, an extremely rich degree of stochasticity is available to ReRAM-based TRNGs to harvest~\cite{ambrogio2014statistical1,gaba2014memristive,guan2012switching}.

\subsection{Random Telegraphic Noise Characteristics}\label{subsub:dependence}
RTS noise is often observed in very small specimen such as microscale pn-juctions and FETs~\cite{realov2010random}, metal contacts, e.g. metal-insulator-metal tunnel junctions~\cite{wang2016relaxation} and nanotubes~\cite{jhang2004random}. Before we introduce RTN harvesting circuitry, it is important to investigate RTS characteristics of our fabricated ReRAMs. 

\begin{figure}[tb]
\centering
{\includegraphics[width=3.1in]{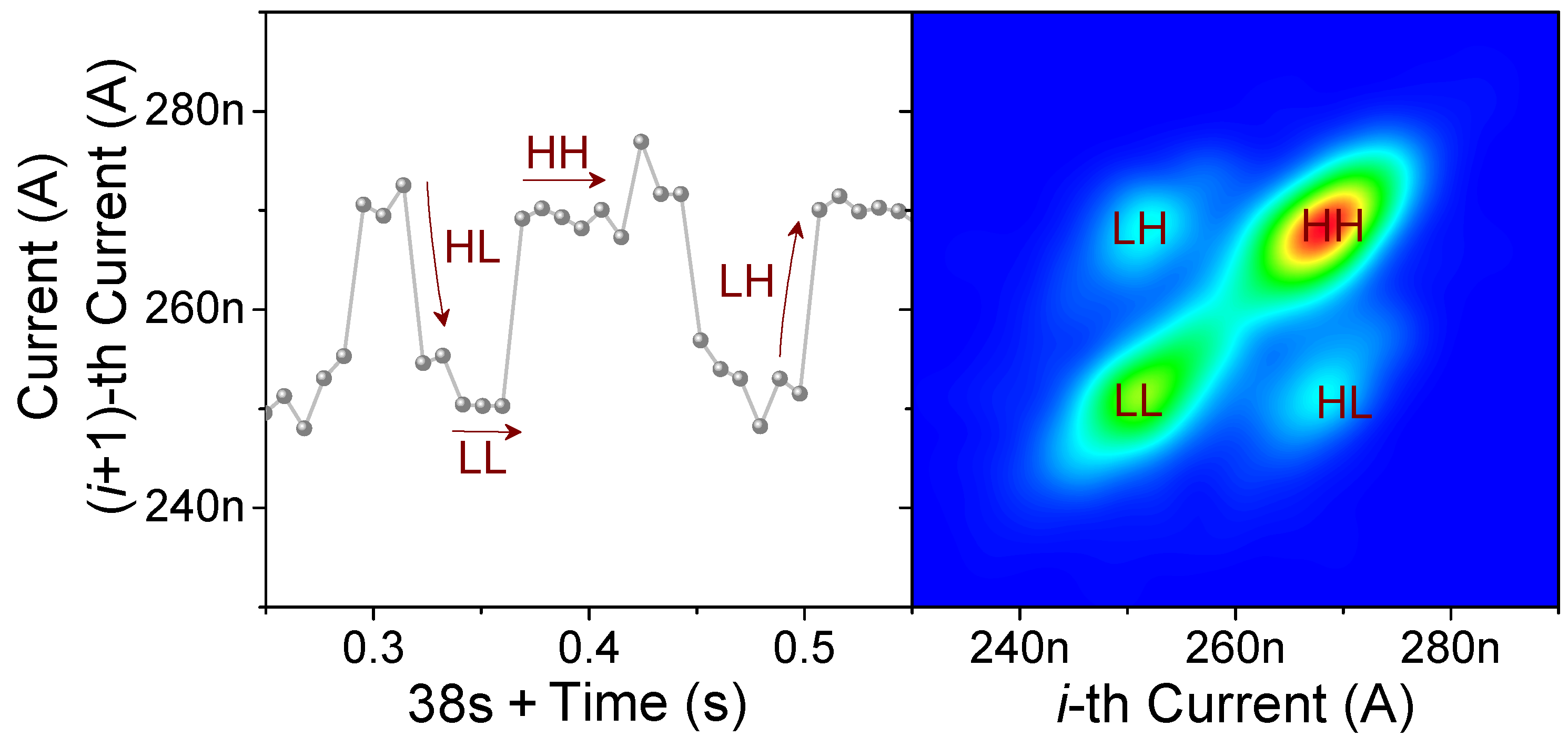}}
\caption{Observed RTS behaviors in a number of $a$-STO-based ReRAMs. A time trace and time-lag plot (TLP) of RTS is presented that shows four distinct transitions. LH and HL represent low-to-high (capture) and high-to-low (emission) level transitions, respectively, which are demonstrations of capture and emission of carriers in a trap/defect. HH and LL are cases that no transition takes place.}\label{fig:TLPsingle}
\end{figure}

We use time leg plots (TLP) to visualize current levels and transition between them~\cite{maestro2016new}, as shown in Fig.~\ref{fig:TLPsingle}, the TLP can be drawn by plotting the RTS data sequence on an {\it x} plane versus a delayed data sequence on an {\it y} plane. In case of a single trap, TLP clearly shows carrier transition from emission to capture (LH) and from capture to emission (HL) in upper-left and bottom-right corners, respectively. We also highlight the other corners as HH and LL for those situation that a carrier stays in captured or the trap stays vacant. The figure shows HH clearly stands out as having the most number of appearance in the acquired data. While switching for capture (LH) and emission (HL) occurs at random times, the balance of color could tell a lot about predictability of RTN-based TRNG which does not have sophisticated post-processing. This important piece of analysis is missing in a number of literature reporting ReRAM\rq{}s RTN-based TRNGs including Refs.~\cite{huang2012contact,chen2016unified}.

Here we not only rely on our data but also reporting numerous experimental evidence reported in Refs.~\cite{balatti2016physical,ambrogio2014statistical2,Noise-induced1,ielmini2010resistance,balatti2015true,ambrogio2014impact} that is difficult to activate and control ReRAM\rq{}s RTN amplitude, average frequency and stability. While these reports could potentially undermine previous ReRAM-based RTN works, they unanimously endorse that amplitude and average frequency of the RTN source cannot be predicted in both high-resistance and low-resistance states (HRS and LRS). It has also been shown that RTN in HRS is activated/deactivated without predictability~\cite{ambrogio2014impact}.

We argue that if proper harvesting technique is used to take advantage of the uncontrollable nature of RTN, RTN could be one of the most true sources of randomness in solid-state devices. Our data, presented in Figs.~\ref{fig:time} and~\ref{fig:TLPsingle} confirms that average frequency of RTN in our devices is uncontrollable and achieving a balanced TLP is extremely difficult, if not impossible. However, as shown in Fig.~\ref{fig:time}, it can be concluded that we have a relatively stable control over RTN amplitude by maintaining a solid control of $V_{\rm READ}$ at the nonlinear HRS curve shown in Fig.~\ref{fig:IV}. As it is expected, studies also suggest strong correlation between low-frequency noise amplitude in oxide-based ReRAMs and resistance values~\cite{fang2013current}. The method for imposing this control using a negative feedback loop is described in Section~\ref{sec:rng-subsec:propcir}.

\subsection{Random Telegraphic Signals and Environmental and Process Factors}\label{subsub:envir}
To get a clear picture of RTS in ReRAM-based TRNGs, we analyze several characteristics of RTS dependence on important factors including a device size, applied potential and temperature.

For scaled FET devices relative RTS amplitude in drain current ($\Delta I_{\rm d}/I$) has shown to decrease by increasing channel area ($W\times L$)~\cite{da2016physics, yasuda2008flicker}. Similarly, area-dependence in ReRAMs is a strong~\cite{fang2013current,ielmini2010resistance}. ReRAMs show increased RTS amplitude at higher resistances and smaller devices~\cite{fang2012area, soni2010probing,song20151}.

\begin{figure}[tb]
\centering
\subfigure[]{\includegraphics[width=2.6in]{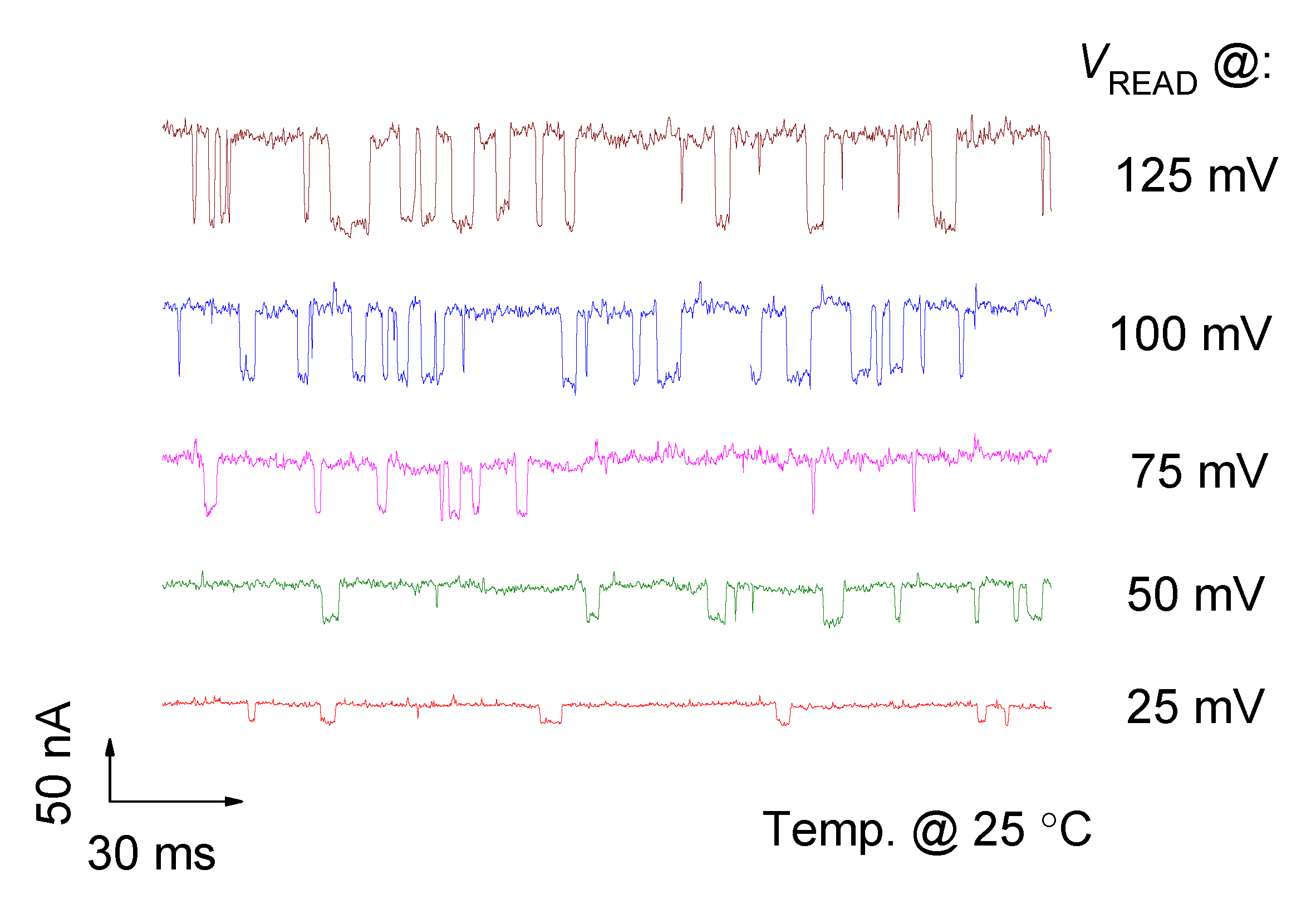}\label{fig:voltdepend}}
\subfigure[]{\includegraphics[width=1.5in]{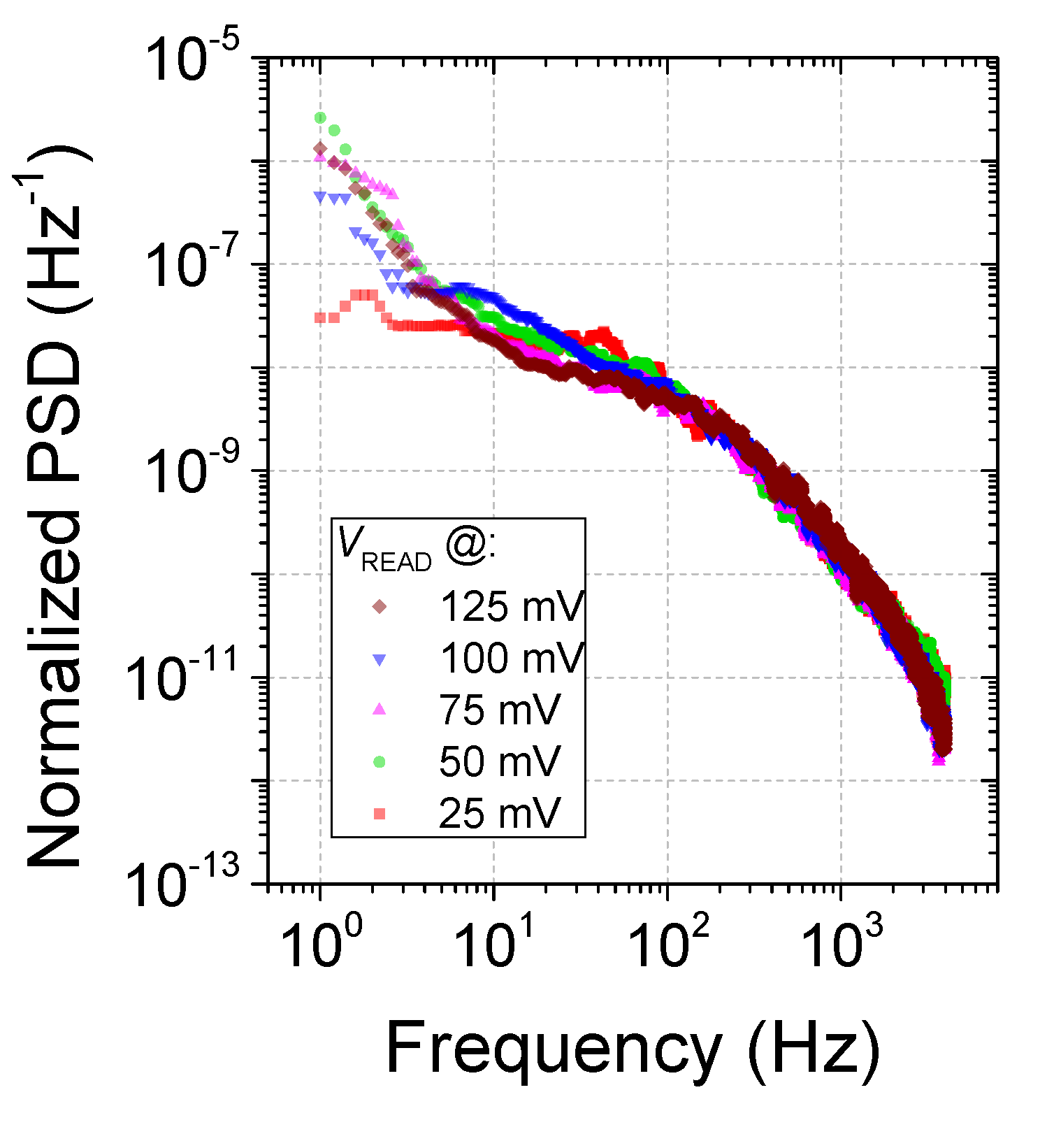}\label{fig:PSD_V}}
\subfigure[]{\includegraphics[width=1.5in]{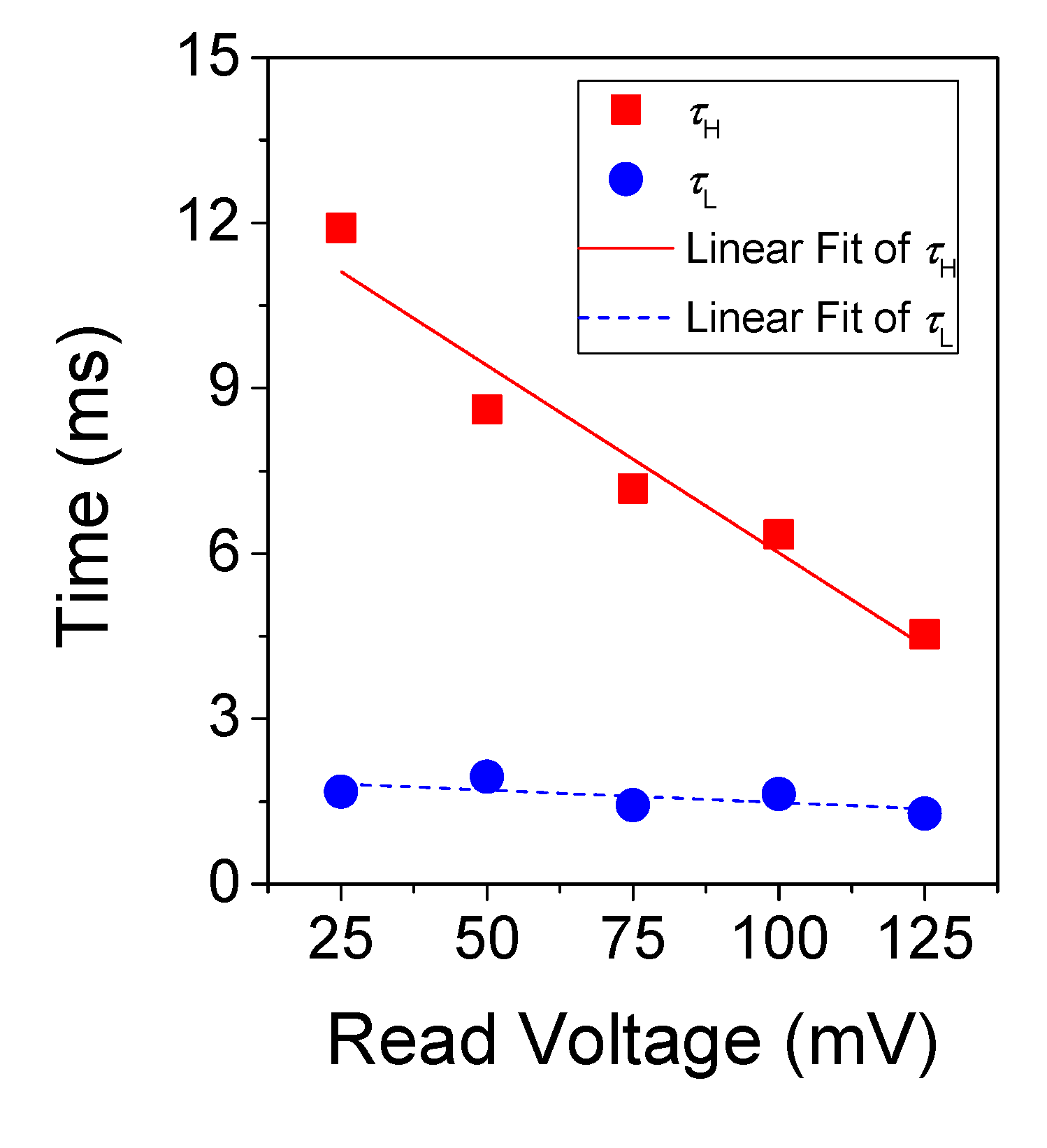}\label{fig:Tau_V}}
\subfigure[]{\includegraphics[width=1.5in]{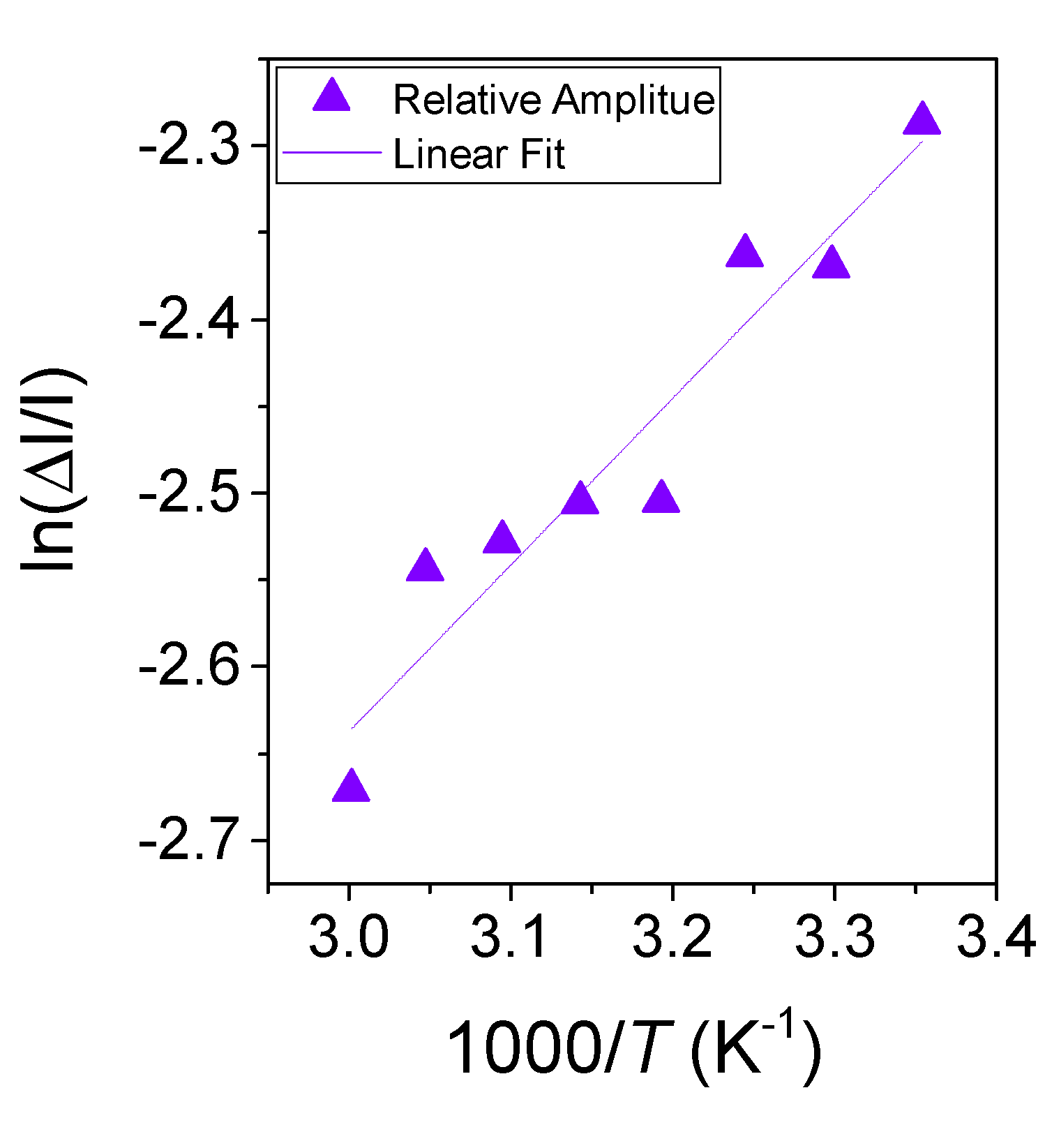}\label{fig:RelAmp_T}}
\subfigure[]{\includegraphics[width=1.5in]{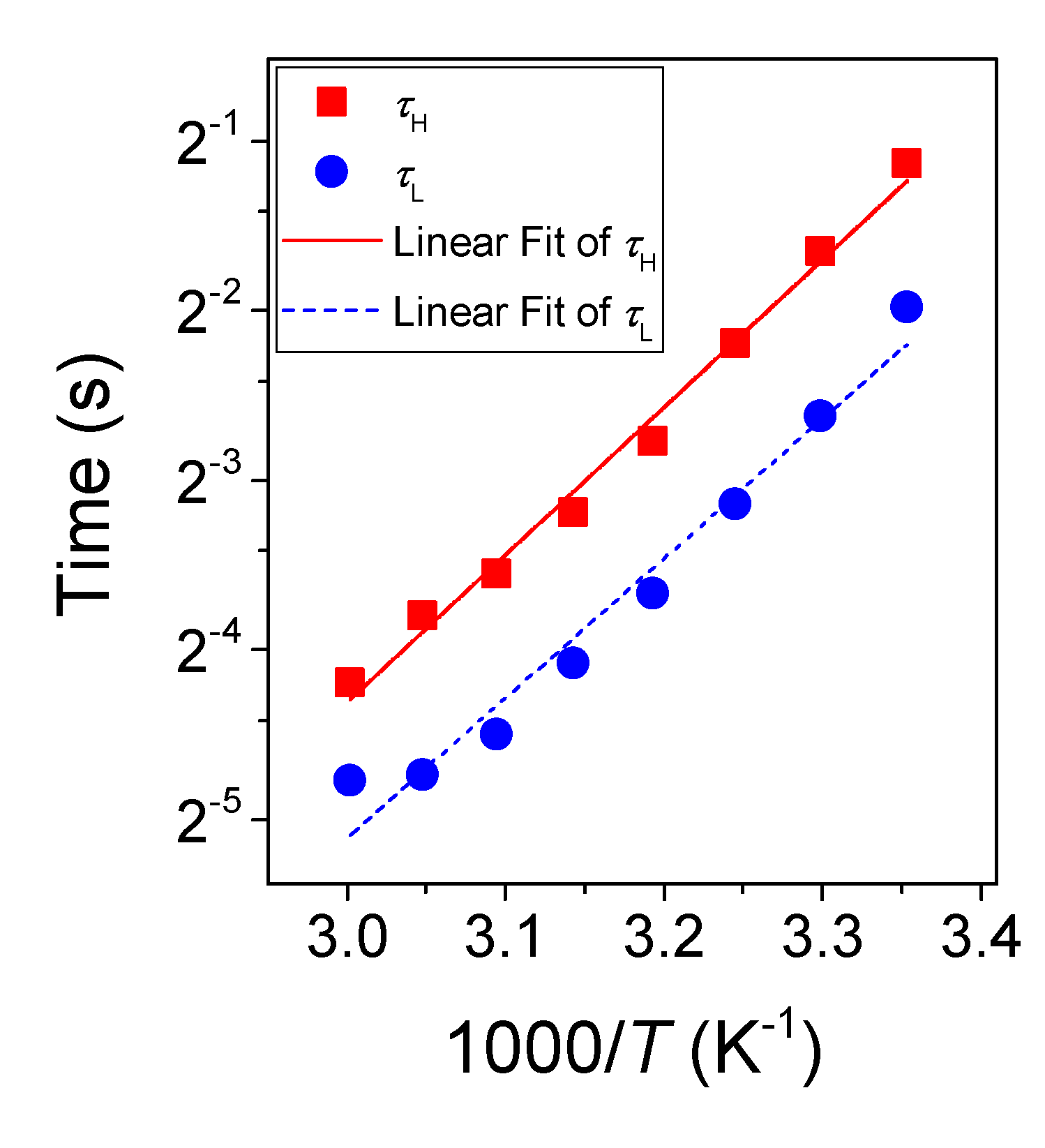}\label{fig:Tau_T}}
\caption{Voltage and Temperature dependence of RTS. (a) Time trace of RTS in $a$-STO-based ReRAM at different $V_{\rm READ}$s. Increased RTS transition rate by rising $V_{\rm READ}$ is observed. (b) Normalized PSD when $V_{\rm READ}$ is swept from $25$~mV to $125$~mV. Normalized PSDs at different $V_{\rm READ}$s follow the similar trend. (c) Average capture, $\tau_{\rm H}$, and emission, $\tau_{\rm L}$, times are shown. $\tau_{\rm H}$ demonstrates significant decrease while $\tau_{\rm L}$ shows almost no voltage dependency. It implies that RTS level transition rate can be adjusted by controlling $V_{\rm READ}$. (d) Relationship between $\Delta I/I$ and temperature. (e) Thermal activation RTS fluctuation represented by average time constants decrement of $\tau_{\rm H}$ and $\tau_{\rm L}$ at higher temperature.}
\end{figure}

In terms of applied potential-dependence of RTS, our measurement suggests two main changes in RTN when $V_{\rm READ}$ is swept. In Fig.~\ref{fig:voltdepend}, we observed the transition rate, corresponding to the average capture and emission times ($\tau_{\rm H}$ and $\tau_{\rm L}$) are dependent on the applied voltage, which in effect means a different conductance point on HRS curve in Fig.~\ref{fig:IV}. The noise power, however, almost follow similar trend at all READ voltages when normalized trend is considered to the $V_{\rm READ}$ (see Fig~\ref{fig:PSD_V}). Normalized PSD is the most important factor as in reality, normalized PSD identifies SNR. The rise in applied voltage and consequently the increase in absolute value of current passing through the device resulted in steady decrease in $\tau_{\rm H}$, capture time, while $\tau_{\rm L}$, emission time, shows a much weaker correlation as shown in Fig.~\ref{fig:Tau_V}.

Temperature dependency of RTS is another important factor. We found that, while RTS amplitude ($\Delta I$) is maintained approximately constant, absolute current ($I$) grows over the temperature range, as expected. This trend shows in Fig.~\ref{fig:RelAmp_T}. As suggested in literature (Ref.~\cite{soni2010probing}), this indicates that RTN is most likely initiated from the same defect(s) during the measurement. We also extracted $\tau_{\rm H}$ and $\tau_{\rm L}$ at different temperatures. Fig~\ref{fig:Tau_T} clearly shows a rapid descent in both $\tau_{\rm H}$ and $\tau_{\rm L}$ as temperature rises, which implies the RTS fluctuation becomes more frequent, yet timing remain stochastic.

\section{$a$-STO-based ReRAM Fabrication}\label{sec:fab}

Before presenting RTN harvesting circuitry and proposed TRNG, it is important to describe our fabrication steps. Using standard photolithography we fabricated a stack of the following material layers. A $20$ nm a Pt and $5$ nm Ti adhesion layers are deposited on a SiO$_2$/Si substrate using electron-beam evaporation to define the bottom electrode (BE). A $22$ nm amorphous SrTiO$_3$ ($a$-STO) thin film was sputtered on top of the BE. Finally, a $20$ nm/$10$ nm of Pt/Ti film was deposited by electron-beam evaporation as top electrode (TE). The whole deposition is completed at room temperature. Our fabricated ReRAMs are attributed to localized accumulation of oxygen vacancies along the defect structure across the device~\cite{niliNano, nili2015donor}. Oxygen vacancy is known to facilitate the formation and rupture of nano-filaments, which is responsible for the bipolar switching between HRS and LRS~\cite{nili2015donor}. Electrical characterization of ReRAM and measurement data was gathered with Keithley 4200 Semiconductor Characterization System (SCS). Full details on the electrical, electroforming, and switching characteristics of the a-STO memristors can be found in Refs.~\cite{niliNano, nili2015donor, nili2016microstructure}.

\section{A ReRAM\rq{}s TRN-based True Random Number Generator}\label{sec:rng}
In this section, we describe our harvesting circuit for a ReRAM-based TRNG for which RTS is used as a source of randomness.

\subsection{Proposed ReRAM-based TRNG}\label{sec:rng-subsec:propcir}
\begin{figure}[!tb]
\centering
\subfigure[]{\includegraphics[width=2.7in]{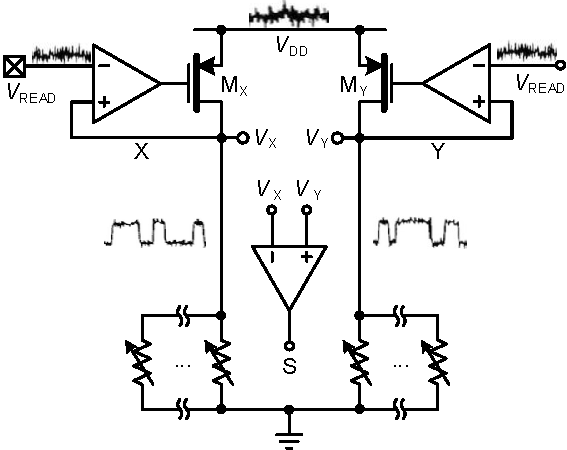}\label{fig:circuit}}
\subfigure[]{\includegraphics[width=2.0in]{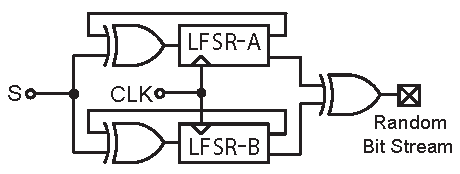}\label{fig:LFSR}}
\caption{The proposed ReRAM-based TRNG circuit. (a)~A differential circuit is proposed that utilizes two amplifiers in negative feedback loop configuration in order to regulate $V_{READ}$ solidly. Identical number of one or more parallel ReRAMs are placed in both branches, which act as sources of RTS. The differential nature of this configuration results in effective rejection of supply voltage and $V_{\rm READ}$ noise, hence, reduced bias in the random output. (b) LFSR-based post-processing circuit is a digital circuitry, which by itself (if output is fed to input) implements a {\it pseudo}-RNG.}
\end{figure}

Fig.~\ref{fig:circuit} presents a differential readout (harvesting) circuit for ReRAM\rq{}s RTSs. Negative feedbacks to amplifiers from nodes X and Y, help to regulate $V_{\rm X}$ and $V_{\rm Y}$ at $V_{\rm READ}$. Loop bandwidth identifies the frequency that $V_{\rm X}$ and $V_{\rm Y}$ are fixed. While our measurement done on a pair  ReRAMs which are placed both sides, it is also possible that ReRAMs are put in parallel, which to some extend helps stability of the loops by reducing overall ReRAM part resistance. The clamping amplifier compares the potential, $V_{\rm X}$ to the applied bias, $V_{\rm READ}$ with a negative feedback loop. Rapid RTN jumps would result sufficient $\Delta V$ at either $V_{\rm X}$ or $V_{\rm Y}$, which means the loop needs some time to settle. Due to random fluctuation on X and Y in time, the proposed TRNG is shown to be capable of generating true randomness in the output according to our NIST and machine learning evaluation. 

The differential nature of this TRNG results in effective supply voltage ($V_{\rm DD}$) and $V_{\rm READ}$ noise rejection. Temperature that is shown to increase RTS activities also influence both ReRAM branches in approximately similar manner, therefore, its impact is significantly suppressed at the circuit level. Radiation attacks or delivering RF energy to the chip would also affect both branches similarly due to small footprint of this TRNG (like many other on-chip differential TRNGs).

Our analysis show, our differential TRNG circuit\rq{}s supply rejection ratio (PSRR) at $100$~Hz is almost an order of magnitude greater than PSRR of its single ended rival presented in Refs.~\cite{huang2012contact,chen2016unified}. $V_{\rm READ}$\rq{}s noise rejection, which is considered common-mode for the whole circuit due to circuit configuration, is also very strong in our implementation. In single ended circuit, a reference voltage is in charge of determining the output bit, which could easily be tampered by application of RF energy, temperature or other forms of delivering energy and/or noise to the system. Our proposed differential circuit take advantages of all other known differential signaling benefits such as higher output swings, simpler biasing and higher linearity. Our analysis also shows offset on the negative feedback loop amplifiers is likely to affect both similarly due to the fact that they are sitting in very close proximity of each other. It is important to note that transmission gate addressing elements are removed in Fig.~\ref{fig:circuit} presentation for the sake of simplicity. 
 
Since the signal is taken differentially, an amplifier that senses differential signals is required at the output. This is shown by the module taking $V_{\rm X}$ and $V_{\rm Y}$ and provides digital signal S at the output, which then is fed to a post-processing unit, shown in Fig.~\ref{fig:LFSR}. We evaluated entropy before and after the post-processing unit. The entropy of bit-stream signal S is $0.97$, which is significantly higher than those of single-ended method at $0.93$. Post-processing improves the entropy to $0.99$, closer to ideal entropy of $1.00$.

\begin{figure}[hb]
\centering
\subfigure[]{\includegraphics[width=3.25in]{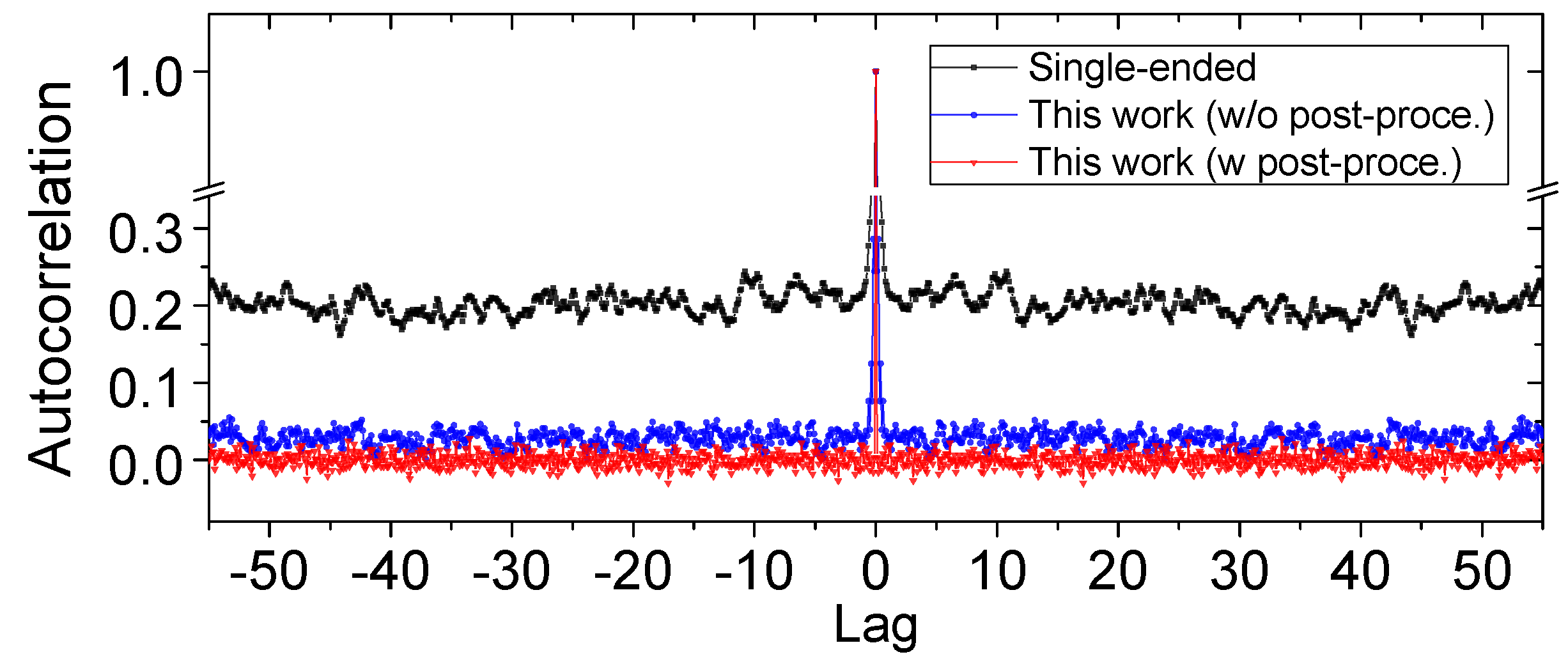}\label{fig:Auto}}
\subfigure[]{\includegraphics[height=1.6in]{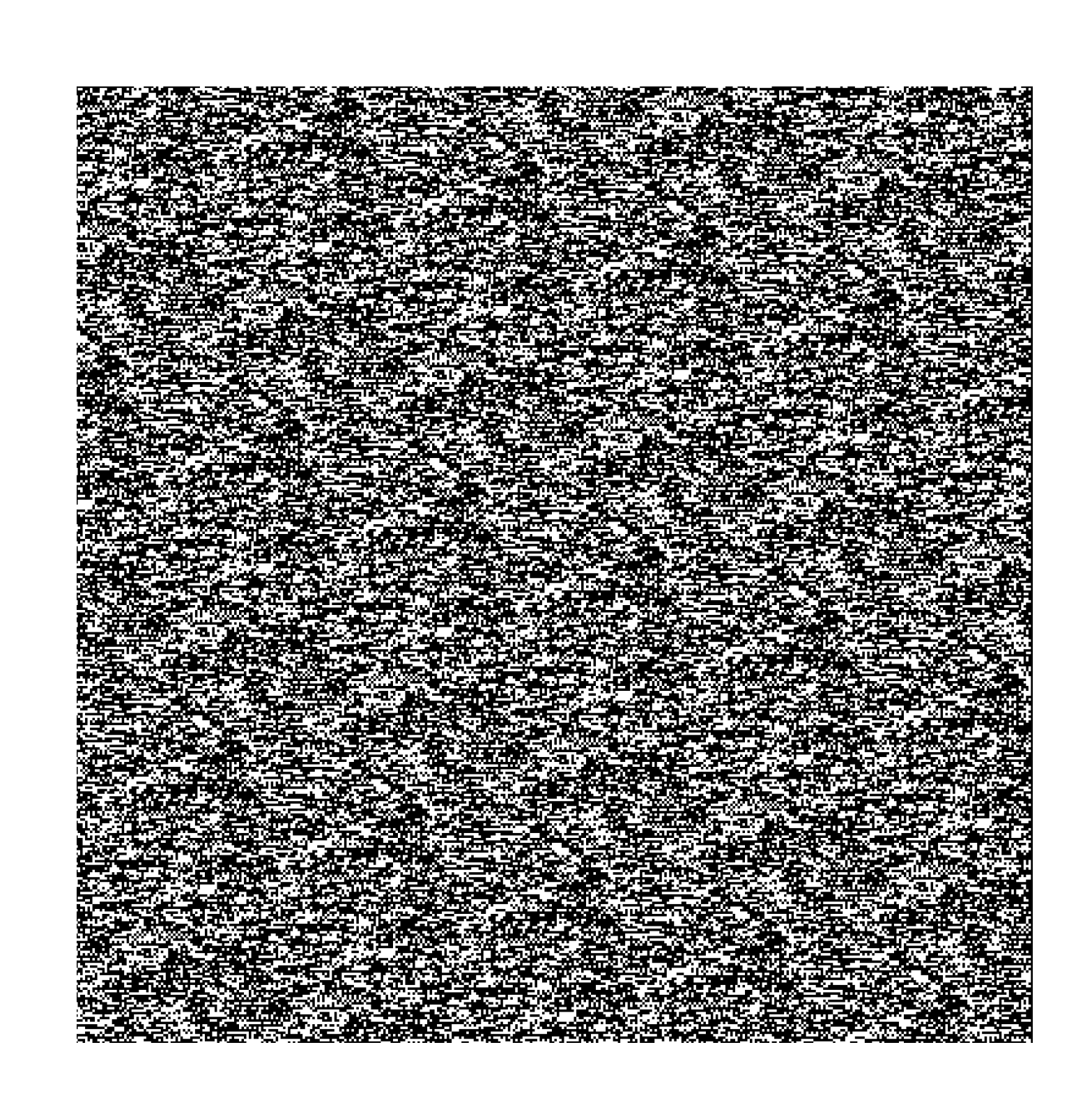}\label{fig:bp_diff}}
\subfigure[]{\includegraphics[height=1.6in]{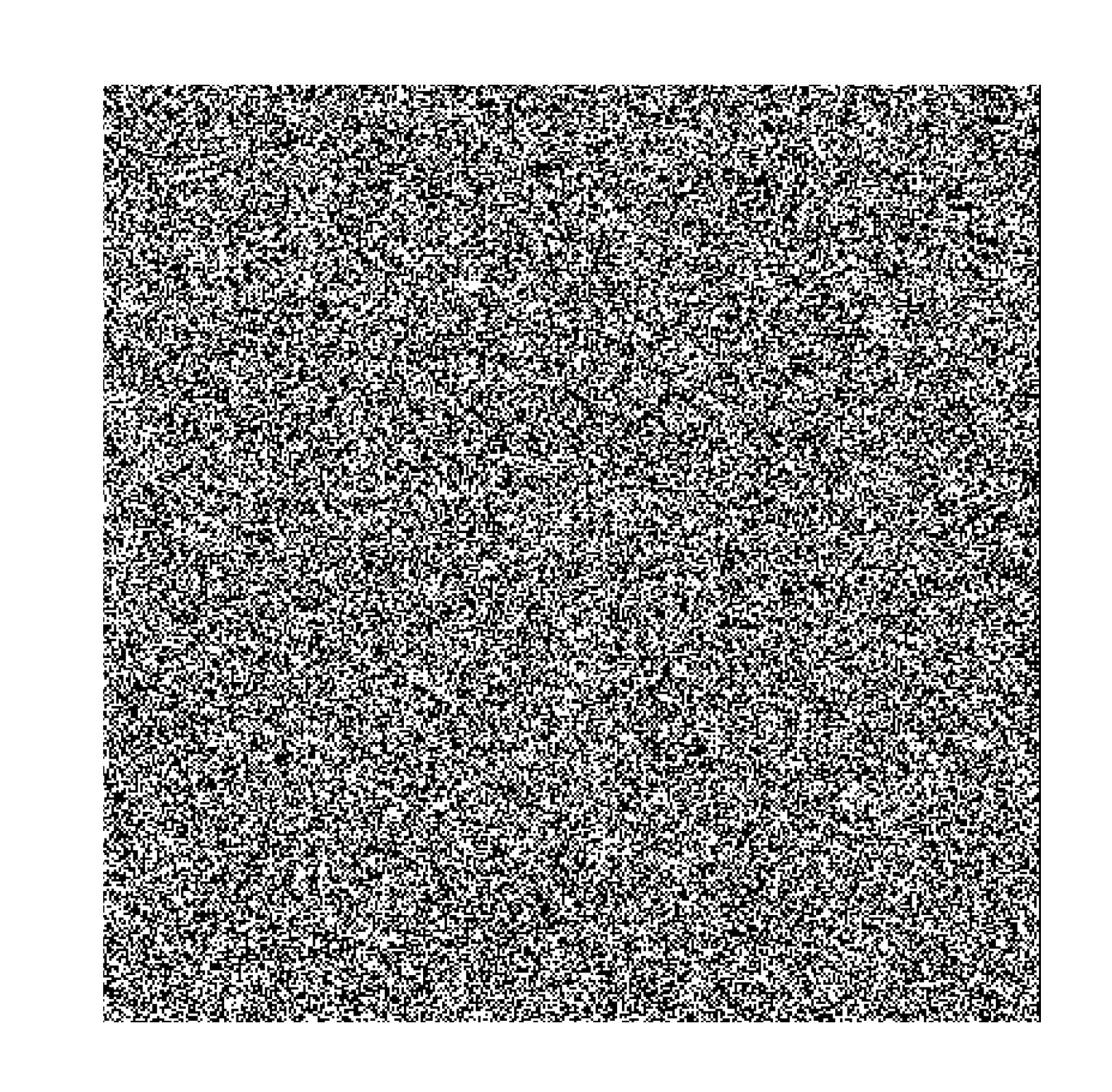}\label{fig:pp_diff}}
\caption{(a) Autocorrelation comparison between single-ended and differential RTN harvesting implementations. The differential method is tested with and without a post-processing unit and the outcome is clearly in favor of differential RTN harvesting technique with post-processing unit. (b) Bitmap generated using a portion of data, more than $2^{17}$ bits, at the output S in Fig~\ref{fig:circuit}. (c) Shows the same number of bits after post-processing (see Fig~\ref{fig:LFSR}).}
\end{figure}

Another way to examine randomness is the autocorrelation. We plot autocorrelations of single-ended implementation and the proposed TRNG with and without post-processing. Results shown in Fig.~\ref{fig:Auto} indicates existence of significantly higher autocorrelation in single-ended RTN harvesting circuits, which could be the outcome of a systematic bias generated as the result of noise on the supply, reference voltage and temperature fluctuations. The sequence generated by our differential method, on the other hand, produces almost no autocorrelation, under identical condition.

The effectiveness of the introduced post-processing unit could be seen in the autocorrelation analysis, where a clear improvement can be observed in Fig.~\ref{fig:Auto}. A way to qualitatively analyze randomness is to visualize some portion of random data as bitmap. Before we report our formal randomness analysis in the next section, Figs.~\ref{fig:bp_diff} and~\ref{fig:pp_diff} could qualitatively represent that no obvious pattern could be observed at the output of post-processing unit.

\section{RNG Randomness Evaluation}\label{sec:eval}
To formally investigate the degree of randomness of our TRNG, entropy analysis and NIST statistical tests are reported in this section. In addition, we include a predictive machine learning evaluation to estimate the level of predictability in our TRNG.

Here, we use statistical test suit developed by the US\rq{}s National Institute of Standard and Technology (NIST) in order to evaluate randomness of our TRNG. The test suite includes a total of 15 different tests with two similar tests running on different directions of bit-stream; hence, 17 tests~\cite{rukhin2001statistical}. Our ReRAM-based TRNG successfully passed all tests with a significance level of $0.01$. Tests, outcome and $p$-values are shown in Table \ref{tab:NIST}. 

We also run prediction test using recurrent neural network (RNN) with long short-term memory (LSTM). Multiple measurements show an average prediction rate between $49.790\%$ and $51.385\%$, which shows unpredictability of our TRNG\rq{}s bit-sequence is close to an ideal level.

\begin{table}[ht]
\centering
\caption{Successful Pass Results of NIST Test on ReRAM-based TRNG.}\label{tab:NIST} 
\begin{threeparttable}
\begin{tabular}[ht]{lccc}
\hline\hline
{\bf Statistical Test} & {\bf Result} & {\bf $p$-value }\\
\hline\hline
Frequency & Pass& 0.370 \\
Block Frequency  &  Pass  & 0.063\\
Cumulative sum (Forward) &  Pass & 0.319 \\
Cumulative sum (Backward)  &  Pass  &0.506 \\
Runs &  Pass  & 0.868\\
Longest Run of 1's & Pass  &  0.246\\
Rank &  Pass &0.779 \\
FFT &  Pass  &0.119 \\
Nonoverlapping Templates &  Pass & 0.011$^\ast$ \\
Overlapping Templates  &  Pass  & 0.597\\
Universal & Pass &0.637\\
Approximate Entropy &  Pass  & 0.846\\
Random Excursions  &  Pass   & 0.069$^\ast$\\
Random Excisions Variant &  Pass & 0.049$^\ast$\\
Serial  &  Pass  &0.417\\
Serial  &  Pass  &0.082\\
Linear Complexity  &  Pass  & 0.787 \\
\hline\hline
\end{tabular}
\vspace*{0.1cm}\scriptsize{$^\ast$Lowest}
\end{threeparttable}
\end{table}

\section{Conclusion}
In this paper, we presented a novel and effective way of generating true random bit sequences by introducing a differential random telegraphic noise harvesting technique. This approach is less sensitive to common-mode noises (e.g, noise on supply and reference voltages) and potentially more immune to temperature disturbances and electromagnetic radiation. We have shown that autocorrelation in reported single-ended RTN readout could increase significantly in presence of common-mode noise and fluctuations of environmental factors. We have reported successful evaluation using standard true randomness tests and the use of advanced deep learning techniques.

\end{document}